\documentstyle[12pt,fleqn,epsf]{article}

\newcommand{\noi}{\noindent}
\newcommand{\ra}{\rightarrow}
\newcommand{\eq}{\begin{equation}}
\newcommand{\en}{\end{equation}}
\newcommand{\eqa}{\begin{eqnarray*}}
\newcommand{\ena}{\end{eqnarray*}}

\newcommand{\caM}{{\cal M}}
\newcommand{\caD}{{\cal D}}
\newcommand{\bpsi}{{\bar \psi}}
\newcommand{\meff}{m_{e\! f\! f}}
\newcommand{\AmS}{{\protect\the\textfont2
  A\kern-.1667em\lower.5ex\hbox{M}\kern-.125emS}}

\begin{document}
\hbox{}\hfill {\small DESY-97-195}\\
\hbox{}\hfill {\small HLRZ1997\_60}\\
\hbox{}\hfill {\small JINR-E2-97-281}\\
\hbox{}\hfill {\small HUB-EP-97/69}

\renewcommand{\thefootnote}{\fnsymbol{footnote}}
\setcounter{footnote}{1}

\begin{center}
\vspace*{1.0cm}

{\LARGE Hadron masses in qQCD with Wilson fermions 
  near the chiral limit\footnote{Contribution to LATTICE 97 --- 
  XVth Int. Symp. on Lattice Field Theory, Edinburgh, Scotland.}}

\vspace*{0.5cm}
{\large
A. Hoferichter$\mbox{}^a$,
E. Laermann$\mbox{}^b$,
V.K. Mitrjushkin$\mbox{}^c$,\\
M. M\"uller-Preussker$\mbox{}^d$,
P. Schmidt$\mbox{}^b$
}
       
\vspace*{0.2cm}
{\normalsize
{$\mbox{}^a$ \em DESY-IfH and HLRZ, Zeuthen, Germany}\\
{$\mbox{}^b$ \em Fakult\"at f\"ur Physik, Universit\"at Bielefeld, Germany}\\
{$\mbox{}^c$ \em Joint Institute for Nuclear Research, Dubna, Russia}\\
{$\mbox{}^d$ \em Institut f\"ur Physik, Humboldt-Universit\"at zu Berlin, Germany}\\
}

\vspace{1cm}
{\bf Abstract}
\end{center}
In quenched lattice QCD with standard Wilson fermions the quark
propagator is computed very close to the chiral limit in the zero-temperature
case. Starting from our experience with lattice QED we employ a modified 
statistical method in order to estimate reliably hadron masses.

\section{Introduction}

Chiral symmetry has remained a problematic topic in lattice gauge theories
over the years. As is well--known, Wilson fermions explicitly violate
the chiral symmetry, which can be hopefully restored in the limit
$\kappa \to \kappa_c(\beta)$ ($\beta$ and $\kappa$ denote 
the gauge coupling and 
hopping parameter, respectively).
The QCD study at the experimentally established
$~m_{\pi}/m_{\rho}~$ value is obstructed by a huge amount of required
computer resources and -- particularly for the quenched approximation --
by large fluctuations in the observables caused by 
near--to--zero eigenvalues of the fermion matrix
(`exceptional configurations'). 

In \cite{qed} a variance reduction technique for estimating  
the pseudoscalar mass $m_{\pi}$ was proposed.
In the case of QED the new estimator for $m_{\pi}$ was proved to work well
close to $\kappa_c$, where the standard estimator fails.
In this work we extend the study to the case of quenched 
lattice QCD.

The Wilson action for lattice QCD is
$S = S_{G} + S_{F}$, where $S_G$ denotes
the plaquette $SU(3)$ gauge action and $S_{F}$   
the fermion part  
\eq
S_{F}(U, {\bar \psi}, \psi) =  \sum_{f=1,2}\sum_{x,y}
\bpsi_x^f \caM_{xy}(U) \psi_y^f~,
\en
with the Wilson matrix $\caM (U)= {\hat 1}-\kappa \caD (U),$
\eq
\caD_{xy} \equiv  \sum_{\mu} \Bigl[ \delta_{y, x+\hat{\mu}} 
P^{-}_{\mu} U_{x \mu}
+ \delta_{y, x-\hat{\mu}} P^{+}_{\mu} U_{x-\hat{\mu},\mu}^{\dagger} \Bigr]
\en
and $P^{\pm}_{\mu}={\hat 1}\pm\gamma_{\mu}$.

The observables of interest are the $\pi$ and $\rho$-meson 
correlators (masses). On every configuration $\{U\}$ we determine
the operators
\eq
\Gamma_{[\pi,\rho]}(\tau) \equiv 
\frac{1}{N_s^6} \sum_{\vec{x},\vec{y}}
\mbox{Tr}
\Bigl( {\cal M}^{-1}_{xy} \gamma_{[5,\mu]} {\cal M}^{-1}_{yx}
\gamma_{[5,\mu]} \Bigr)
\en
\noi where $\tau = |x_4 - y_4|$.

All simulations were performed on a $16^3\times 32$ lattice at
$\beta=6.0$.  The number of the gauge field configurations is $O(100)$.
$\kappa$-values were chosen between $0.1558$ and $0.1570$.  
The inversion of $\cal{M}$ was done with BiCGstabI with restarting.

\section{Correlators and Masses}

%
%
%
%
\begin{figure}[htb]
\vspace{-1.2cm}

\epsfxsize=15.0cm
\epsfysize=13.5cm 
\epsfbox{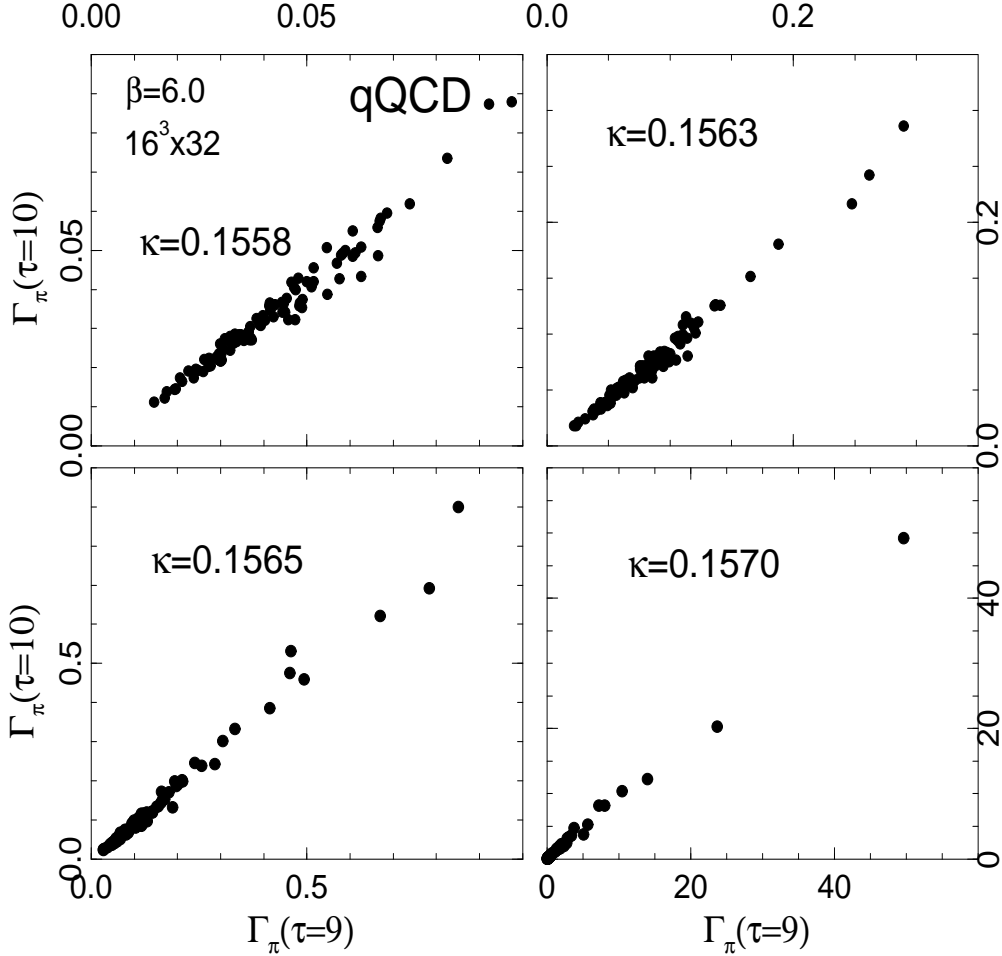}
\vspace{-0.5cm}
\caption{Scatter plots for the pion correlator.
}
\label{fig:scorr}
\end{figure}
%
%

%
%
%
\begin{figure}[htb]
\vspace{-10mm}
\epsfysize=12.0cm
\epsfxsize=14.0cm
\epsfbox{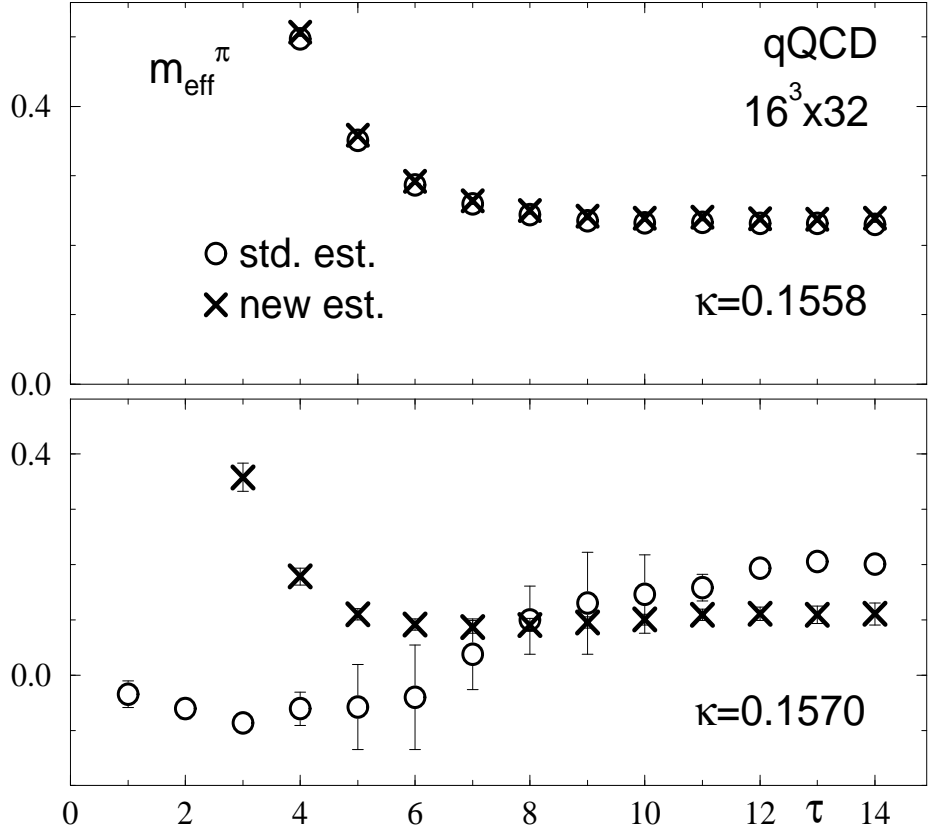}
\vspace{-0.7cm}
\caption{$\meff^{\pi}$ over $\tau$ for two values of $\kappa$. 
(o) denote the standard and (x) the improved estimator.}
\label{fig:tdep}
\end{figure}

%
\begin{figure}[htb]
\vspace{-1.5cm}
\epsfysize=12.6cm
\epsfxsize=14.5cm
\epsfbox{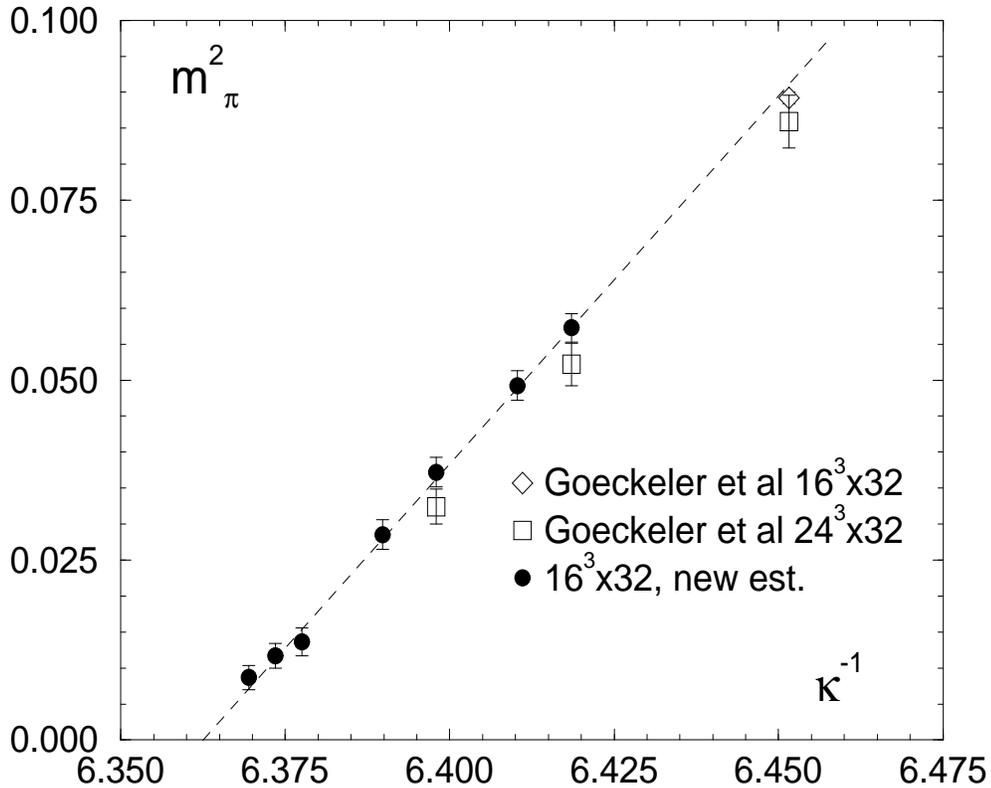}
\vspace{-0.5cm}
\caption{$m_{\pi}^2$ as a function of $~1/\kappa$.}
\label{fig:globalpi}
\end{figure}
%

%
\begin{figure}[htb]
\vspace{-1.5cm}
\epsfysize=13.6cm
\epsfxsize=15.0cm
\epsfbox{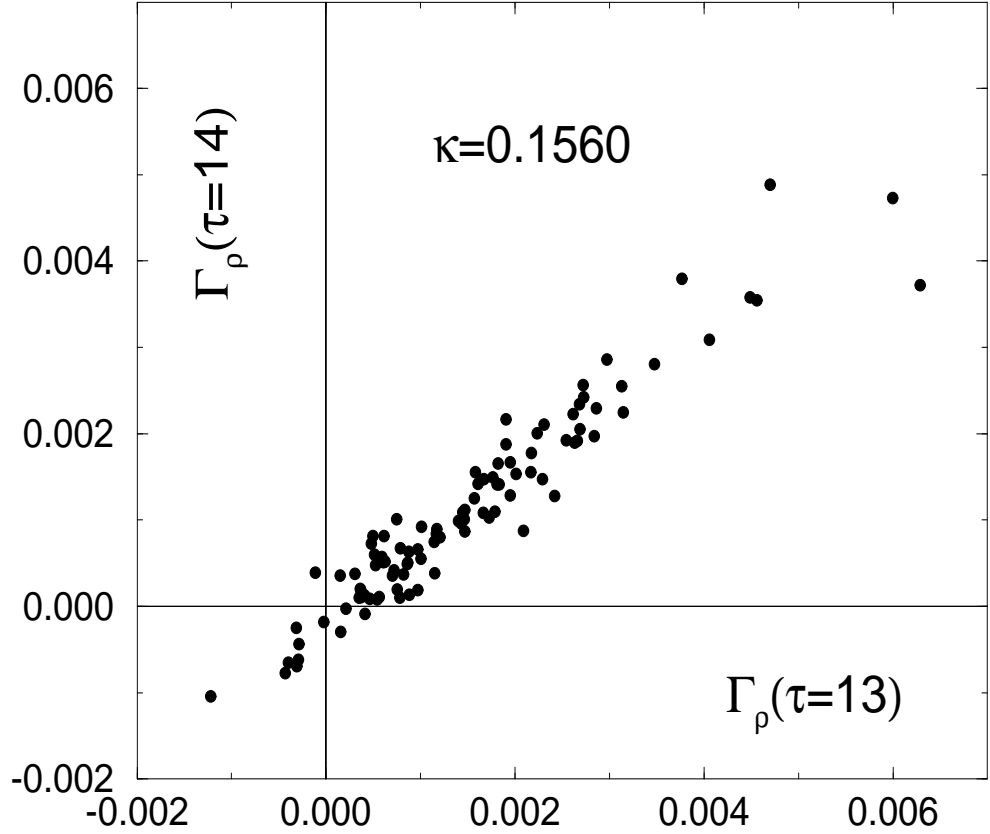}
\vspace{-0.5cm}
\caption{Scatter plot for $\rho$--meson correlator.}
\label{fig:rho}
\end{figure}

The standard definition of the effective mass $\meff(\tau)$
is given by the equation
\eq
\frac{\cosh (\meff (\tau )(\tau +1-\frac{L_4}{2}))}
{\cosh (\meff (\tau )(\tau -\frac{L_4}{2}))}
= \frac{\langle\Gamma_{[\pi,\rho]}(\tau +1)\rangle}{\langle
              \Gamma_{[\pi,\rho]}(\tau )\rangle}~.
              \label{m_eff}
\en
The plateau in the $\tau$--dependence of $\meff(\tau)$ is supposed to 
define the `true' mass $m_{[\pi,\rho]}$.
However, for $\kappa \rightarrow \kappa_c(\beta)$ the values of 
$\Gamma_{\pi}(\tau)$ become strongly fluctuating. 
There is no reliable estimate for the averages 
$\langle \Gamma_{\pi}(\tau)\rangle$,  
the approach to $\kappa_c(\beta)$ seems extremely difficult \cite{eig1,eig2}.

In \cite{qed} it was shown that a new estimator for the pseudoscalar
mass can be defined which is identical to the standard estimator in the case
of linear correlations
\eq
\overline{y}(x) = {\rm C}\cdot x,
\label{corr}
\en
where $x\equiv\Gamma_{\pi}(\tau)$; $y\equiv\Gamma_{\pi}(\tau+k); ~k>0$,
and $\Gamma_{\pi}(\tau+k)$, $\Gamma_{\pi}(\tau)$
are calculated on individual configurations. 
$\overline{y}(x)$ stands for
the conditional average of $y(x)$ at a fixed value of $x$. 
If (\ref{corr}) holds with $~x \neq 0~$, then
\eq
\frac{\langle \Gamma_{\pi}(\tau + k) \rangle }{\langle \Gamma_{\pi}(\tau) 
\rangle} \equiv \frac{\langle y \rangle }{\langle x \rangle}
= \left\langle \frac{y}{x} \right\rangle
\equiv \left\langle \frac{\Gamma_{\pi}(\tau + k)}{\Gamma_{\pi}(\tau)} 
\right\rangle~,
\label{xy}
\en
and the effective masses are determined by the r.h.s. of (\ref{xy}).
The important observation \cite{qed} is that the ratio
$\Gamma_{\pi}(\tau + k)/\Gamma_{\pi}(\tau)$ calculated on individual
configurations does not suffer from near--to--zero eigenvalues.
The average of this ratio is statistically well-behaved, in
contrast to the ratio of averages.

A typical situation is shown in Fig.\ref{fig:scorr}
where $\Gamma_{\pi}(\tau+1)$ over $\Gamma_{\pi}(\tau)$ is depicted 
for $\tau=9$ at several $\kappa$'s (note the different scales).
We do not find any qualitative difference to the quenched
compact QED$_{\rm 4}$ case. Therefore, we expect the improved estimator 
to be applicable also for quenched QCD.

Fig.\ref{fig:tdep} represents the effective mass $\meff^{\pi}(\tau)$
for two different values of $\kappa$. For the smaller $\kappa$ we 
obtain very good agreement between the two estimators, which with more 
statistics, should become identical.
In the lower plot, near to $~\kappa_c$, the standard estimator 
and its error bars
cannot be trusted anymore. In contrast, the improved estimator
gives a very nice plateau in $\meff^{\pi}$ at $\tau > 5$ 
which still is statistically well-defined.

The extrapolation $m_{\pi} \ra 0$ is shown in Fig.\ref{fig:globalpi}. 
The PCAC--like dependence
\eq
m^2_{\pi} \sim m_q = \frac{1}{2}\left(\frac{1}{\kappa}-\frac{1}{\kappa_c}\right)
              \label{pcac_like}
\en
\noi is nicely fulfilled.
In this figure we included also data from the QCDSF collaboration 
on $16^3\times32$ and $24^3\times32$ lattices \cite{qcdsf} for comparison. 
The data look well-consistent for coinciding lattice sizes.

The $\rho$--meson mass $m_{\rho}$ should not disappear in the limit
$\kappa\to\kappa_c$.  Therefore, the $\rho$--meson correlators
$\Gamma_{\rho}(\tau)$ are expected to be less sensitive with respect 
to near--to--zero eigenvalues.
The standard estimator for $m_{\rho}$ works better, in
contrast to the pion case.
However, it is interesting to note that scatter plots for the 
$\rho$--meson correlators with different $\tau$ show again the linear
correlations between different $\tau$--slices (see Fig. \ref{fig:rho}).

\section{Summary}

We observed that due to strong linear correlations between 
the values of the pseudoscalar correlators at different $\tau$ the ratios
$\Gamma_{\pi}(\tau +1)/\Gamma_{\pi}(\tau)$ do not suffer
from near--to--zero (exceptional) eigenmodes of the fermionic matrix.
This means that for every given configuration all `divergent'
contributions to the correlators are factorized. 

Making use of this observation we proposed another estimator of the
pseudoscalar mass, which is well-defined near 
$\kappa_c(\beta)$ in contrast to the standard estimator.

For $\kappa$--values sufficiently below $\kappa_c(\beta)$, i.e., where the
standard estimator can be reliably defined, both
estimators are shown to be in a very good agreement. 
By approaching $\kappa_c(\beta)$ we observe a PCAC--like dependence 
as in eq.~(\ref{pcac_like}).
For values of $\kappa$ very close to $\kappa_c(\beta)$
the standard estimator fails to work, while the improved one
still fits the same straight line having a very small statistical error.

The new estimator permits to approach the chiral limit
much closer than the standard one would allow.  Especially, systematic
errors induced by the finite volume and/or by the quenched approximation
can be investigated in more detail within the
`critical' region.  In particular, the `critical' value
$~\kappa_c(\beta)~$ can be determined with higher accuracy.

An alternative way to cure the problem of `exceptional configurations'
was proposed in Ref. \cite{eichten}.

\vspace{0.3cm}
\noi {\bf Acknowledgement}

We thank QCDSF (R. Horsley) for providing us with their data for  
$\Gamma_{\pi}(\tau)$.

\end{document}